# Multi-modal MRI sensitive to age:
# Focus on early brain development in infants

Jessica DUBOIS, PhD

*Affiliations:*
INSERM, University of Paris, NeuroDiderot unit UMR1141, Paris, France
CEA, NeuroSpin, UNIACT, Paris-Saclay University, Gif-sur-Yvette, France

*Corresponding address:*
Jessica Dubois
CEA/SAC/NeuroSpin
Bât 145, point courrier 156
91191 Gif-sur-Yvette, France
Email: jessica.dubois@centraliens.net

## Abstract

Exploring the developing brain is a major issue in understanding what enables children to acquire amazing abilities, and how early disruptions can lead to a wide range of neurodevelopmental disorders. MRI plays a key role here by providing a non-invasive way to link brain and behavioral changes. Several modalities are used in newborns and infants to characterize the properties of the developing brain, from growth, morphology to microstructure and functional specialization. Recent multi-modal studies have sought to couple complementary MRI markers to provide a more integrated view of brain development. In this chapter, we describe successively how these approaches have made it possible to assess the early maturation of brain tissues, to link different aspects of structural development, and to compare structural and functional brain development.

## Key words

Maturation of grey and white matter, premature newborns, typical infants, structural MRI, anatomical MRI, diffusion MRI, relaxometry MRI, functional MRI, EEG/MEG, quantitative parameters.

2## Abbreviations

| | |
|---|---|
| AD | axial diffusivity |
| ASL | arterial spin labeling |
| BOLD | blood oxygen level dependent |
| CHARMED | composite hindered and restricted model of diffusion |
| DARTEL | diffeomorphic anatomical registration using exponentiated Lie algebra |
| DIAMOND | distribution of 3D anisotropic microstructural environments in diffusion-compartment imaging |
| DISCO | diffeomorphic sulcal-based cortical registration |
| DKI | diffusion kurtosis imaging |
| DTI | diffusion tensor imaging |
| EEG | electroencephalography |
| fMRI | functional MRI |
| GMM | gaussian mixture model |
| MD | mean diffusivity |
| MEG | magnetoencephalography |
| MRI | magnetic resonance imaging |
| MTI | magnetization transfer imaging |
| MTR | magnetization transfer ratio |
| NODDI | neurite orientation dispersion and density imaging |
| pCASL | pseudo-continuous ASL |
| PC-MRA | phase-contrast MR angiography |
| QSM | quantitative susceptibility mapping |
| RD | radial diffusivity |
| rs-fMRI | resting-state fMRI |
| SPANGY | spectral analysis of gyrification |
| T1 | longitudinal relaxation time |
| T2 | transverse relaxation times |
| T1/2w | T1/2-weighted |
| w PMA | weeks of post-menstrual age |

# Introduction

During the last trimester of pregnancy and the first post-natal year, the brain grows dramatically and matures intensely (Kostović, Sedmak et al. 2019), allowing newborns and infants to develop sophisticated perceptions of their environment and to acquire complex motor and cognitive skills according to the multiple experiences and learnings they are exposed to (Dehaene-Lambertz and Spelke 2015). Early alterations in normal neural development can lead to a variety of behavioral disorders that manifest in childhood. It is therefore important to gain a better understanding of these phenomena, which requires linking brain and behavioral changes *in vivo*. This is why the use of non-invasive neuroimaging methods such as 'magnetic resonance imaging' (MRI) has revolutionized our knowledge in this field over the last twenty years.

Before describing these recent advances, it is important to remember that MRI measurements only indirectly reflect the complex series of dynamic and intertwined mechanisms observed throughout development at the molecular, cellular and network levels, which occur within a highly constrained but constantly evolving context (Stiles and Jernigan 2010, Kostović, Sedmak et al. 2019). The main lines of cerebral architecture are set up early *in utero*, and then prolonged maturation phenomena are observed, with the growth of connections between neurons, synaptogenesis and pruning mechanisms, myelination, neurochemical maturation, etc. (Dubois, Alison et al. 2020). These changes in the brain tissue microstructure lead to intense macroscopic changes, with exponential increases in brain size and surface, and increasing morphological complexity with the formation of gyri and the folding of primary, secondary and tertiary sulci (Dubois and Dehaene-Lambertz 2015). An elaborate structural and functional connectivity gradually develops between brain regions, relying on both long- and short-range connections (Dubois, Kostovic et al. 2015, Kostović, Sedmak et al. 2019). In addition to increasing the conduction speed of nerve fibers, the process of myelination may play a role in stabilizing connections that are functionally relevant. Most of these complex mechanisms occur over different time periods and at different rates in different in different regions of the brain and functional networks (Flechsig 1920, Yakovlev and Lecours 1967, Kostović, Sedmak et al. 2019), with the primary, unimodal and transmodal associative ones showing staggered trajectories of maturation.

Different MRI modalities make it possible to explore some of these mechanisms, with varying precision and specificity depending on the method. This requires addressing the main methodological challenges posed by MRI of newborns and infants: sensitivity to movements, scanner noise, size of the brain structures in relation to the spatial resolution of images, incomplete tissue maturation involving specific MR characteristics and contrasts, all of which lead to the implementation of dedicated image acquisition sequences and post-processing tools (Dubois, Alison et al. 2020).

In recent years, several groups of researchers and clinicians have turned to so-called "multi-modal" approaches, aimed at coupling several imaging modalities, in order to improve the sensitivity and specificity of MRI in neonates or infants for the integrated study of brain development. With this in mind, this chapter is organized in four parts. First, we present a summary of the main MRI modalities and developmental changes observed in "univariate" studies of the baby brain (I), with more details in the chapters devoted to each of these modalities. We then describe the multi-modal studies performed to assess the early maturation of brain tissues (II), to relate complementary aspects of structural development (II), and finally to compare the structural and functional brain development (IV). In addition to MRI, this last part includes studies that have used 'electro- and magnetoencephalography' (EEG/MEG) to provide complementary measures of neurophysiological maturation.

# I. Overview of age-related changes in MRI modalities

## 1. Anatomical and relaxometry MRI

In anatomical MRI, the signal intensities of brain images are weighted by longitudinal or transverse relaxation times (T1, T2) that depend on water and fat contents and decrease with maturation processes (Dubois, Dehaene-Lambertz et al. 2014, Dubois, Alison et al. 2020). Contrasts between brain tissues thus evolve during development, and successive stages are generally described during infancy since T1 and T2 times decrease more strongly in white matter than in grey matter because of myelination processes. Changes in contrasts are observed on T1-weighted (T1w) images before T2w images because the change in water molecules compartmentalization (leading mostly to T1 shortening during the pre-myelinating state) occurs before the increase of protein and lipid contents with the chemical maturation of the myelin sheath (leading mostly to T2 shortening). To compare the maturation between brain regions or between infants, measuring T1 and T2 relaxation times quantitatively is a more relevant approach than the observation of T1w and T2w images.

## 2. Diffusion MRI

Because of the high water content and the low myelination, the diffusion properties are also very different in the brains of neonates and infants compared with children and adults (Dubois, Dehaene-Lambertz et al. 2014, Dubois, Alison et al. 2020). During the preterm period, different groups have observed an early anisotropy in cortical grey matter and a radial orientation of the main tensor eigenvector as measured with 'diffusion tensor imaging' (DTI) (Ouyang, Dubois et al. 2019), which possibly relies on the early presence of radial glia fibers and apical dendrites of pyramidal neurons. Subsequently, diffusion within the cortical plate becomes isotropic with elongation and complex branches of neural connections (e.g. basal dendrites of pyramidal neurons, thalamocortical fibers). This decrease of DTI anisotropy in the cortex appears to stabilize around term equivalent age whereas diffusivity indices continue to decrease, due to several competitive microstructural mechanisms (Kostović, Sedmak et al. 2019). To better disentangle the progression of these mechanisms during development, recent studies have highlighted the potential of more complex diffusion models such as 'diffusion kurtosis imaging' (DKI) and 'neurite orientation dispersion and density imaging' (NODDI) (Batalle, O'Muircheartaigh et al. 2019, Ouyang, Jeon et al. 2019, Dubois, Alison et al. 2020).

Diffusion MRI parameters also evolve in the developing white matter, as axonal fibers gradually become mature and functional through the process of myelination (Dubois, Dehaene-Lambertz et al. 2014, Dubois, Adibpour et al. 2016, Dubois, Alison et al. 2020). During the preterm period, DTI diffusivities decrease while anisotropy increases in most white matter regions except at cross-roads locations (Nossin-Manor, Card et al. 2013, Kersbergen, Leemans et al. 2014). During infancy and childhood, 'radial diffusivity' (RD) decreases more than 'axial diffusivity' (AD) in bundles, leading to anisotropy increase. Diffusion parameters might be sensitive to the proliferation of glial cell bodies, the extension of oligodendroglial processes, and their wrapping around axons.

## 3. Other quantitative MRI methods

Several other MRI parameters can be measured to quantify the brain maturation (Dubois, Alison et al. 2020). The 'magnetization transfer ratio' (MTR) is thought to reflect mostly the myelin amount as it increases from birth to 2 years of age, at different rates, in the main white matter regions and in the central grey matter nuclei. Nevertheless, in the preterm brain, this technique appears to be sensitive not only to myelin-associated macromolecules, but also to the macromolecular density of axonal cytoskeleton components such as microtubules and neurofilaments (Nossin-Manor, Card et al. 2013). 'Quantitative susceptibility mapping' (QSM) provides information on iron, myelin and macromolecular contents. It shows age-related increase in susceptibility in deep grey matter nuclei in infants, with differences between nuclei (Ning, Liu et al. 2019), and decrease in white matter bundles following a posterior-anterior spatial and temporal pattern (Zhang, Shi et al. 2019). Finally, 'phase-contrast MR angiography' (PC-MRA), 'arterial spin labeling' (ASL) and 'pseudo-continuous ASL' (pCASL) have been used to estimate the cerebral blood flow in the developing brain (Ouyang, Liu et al. 2017, Dubois, Alison et al. 2020), showing age-related increase that reflects brain maturation.

### 4. Multi-compartmental methods

Recently, different modelling approaches based on relaxometry or diffusion MRI data have been proposed to better characterize the brain tissue microstructure by distinguishing different pools of water molecules inside each voxel, based on their distinct characteristics (Dubois, Alison et al. 2020). Analyses of multi-component relaxation assume that at least two or three pools of water contribute to the MR signal sensitized to T1 and T2 relaxometries (Spader, Ellermeier et al. 2013), including water located within the myelin sheath (with relatively short T1 and T2 relaxation times). This provides an estimation of the volume fraction of water related to myelin, which drastically increases with age in the white matter. On the other hand, different multi-component approaches (e.g. 'composite hindered and restricted model of diffusion' CHARMED (Assaf and Basser 2005) or NODDI (Zhang, Schneider et al. 2012)) have been proposed to analyze the MR diffusion signal acquired with multiple shells (i.e. sensitized to multiple b-values) and with multiple gradient directions. The provided parameters also show intense age-related changes during infancy, both in the grey and white matter (Ouyang, Dubois et al. 2019, Dubois, Alison et al. 2020).

### 5. Functional MRI

Brain responses detected with task-based 'functional MRI' (fMRI) also evolve throughout development, because of the progressive specialization and of physiological maturation processes leading to changes in cerebral blood flow, in oxygen consumption, and thus in the 'blood oxygen level dependent' (BOLD) response characteristics (i.e. time-course, amplitude) (Dubois, Alison et al. 2020). Estimating the location of activated regions in a reliable way in newborns therefore requires measuring an accurate model of the hemodynamic response function, in a region-specific manner as the neurovascular coupling might differ between functional systems that are at different stages of maturation. 'Resting-state fMRI' (rs-fMRI) studies have also shown some convergences but also some specific developmental patterns in the early architecture of brain networks (Dubois, Alison et al. 2020).

### 6. Methodological considerations for multi-modal MRI

Therefore, the parameters measured by the different MRI modalities provide different and complementary information on the multiple maturation mechanisms that take place during early brain development. Nevertheless, it is important to note that MRI parameters vary across brains regions including in the adult brain, depending on tissue microstructure: the clearest example is in the white matter where the geometry, compactness and crossings of fibers have a considerable impact on diffusion parameters. The study of the inter-regional variability of maturation therefore requires dissociating it from the variability related to microstructure, which is observed throughout development and at the mature stage, in a more or less consistent manner. This can be taken into account by considering developmental trajectories to assess the maturation asymptotes, or by normalizing infant measurements with adult references (Dubois, Dehaene-Lambertz et al. 2014), making it possible to differentiate maturation and microstructure patterns.

Coupling MRI measurements seems an interesting approach to provide an integrative view of brain development during infancy. However, correlating parameters that vary over the course of development requires considering this dependence on the age of the babies. This can for example be done with a partial correlation method. The first pan of research we aim to describe in the chapter is the use of multi-modal MRI to attempt to quantify the brain tissue maturation more reliably than univariate approaches.

## II.  Multi-modal MRI to assess the early maturation of brain tissues

### 1. *Differentiating brain tissues during early development*

In neonates and infants before ~6-8 months of age, T2w images are generally preferred for delineating the immature white matter, the cortical and sub-cortical grey matter, whereas T1w images are useful for distinguishing the myelinated white matter. Therefore, multi-modal approaches have proposed the combined use of T1w and T2w images to better distinguish between brain tissues and between regions with different maturation, and thus to provide more precise segmentations. But so far the benefit in infants does not seem so obvious (Dean, Planalp et al. 2018).

Based on T1w and T2w images of a large cohort of premature and full-term neonates, a multi-output Bayesian regression technique has been used to model the typical changes in brain contrasts and local tissue shape, according to age at scan, degree of prematurity and sex (O'Muircheartaigh, Robinson et al. 2020). And then it was able to detect focal white matter injury by calculating voxel-wise deviations of a neonate's observed MRI from that predicted by the model, suggesting a clear potential for clinical use.

In the continuity of recent studies conducted in adults, the T1w/T2w ratio has also been proposed as a marker of myelination. In preterm newborns between 36 and 44 weeks of post-menstrual age (w PMA), this measure increases in most cortical regions, suggesting intense maturation and differences between sensorimotor and associative regions (Bozek, Makropoulos et al. 2018). This approach has also allowed to improve the contrast of early myelinating white matter structures (e.g. posterior limb of the internal capsule, corticospinal tract, optic radiations) in neonates (Soun, Liu et al. 2017).

### 2. *Quantifying grey matter maturation*

Multi-parametric MRI approaches have been proposed recently to quantify the maturation of cortical and subcortical grey matter by integrating the complementary information provided by parameters from relaxometry MRI, diffusion MRI, and/or multi-compartment methods. For instance, the microstructural properties of the cortex have been evaluated in preterm infants scanned at term equivalent age (Friedrichs-Maeder, Griffa et al. 2017), showing that values of T1 and DTI 'mean diffusivity' (MD) were lower in the primary sensorimotor cortices than in secondary processing areas, which in turn were lower than in higher-order tertiary areas. This suggested differential patterns of microstructural maturation between functional regions. In typical infants between 1 and 5 months of age, we have mapped the maturational progression of the developing cortex without any hypothesis *a priori* on their anatomical location (Lebenberg, Mangin et al. 2019), with the goal to consider jointly changes related to various maturation mechanisms, including changes in cell and membrane density, in water and iron content, in relation with the development of dendritic arborization, synaptogenesis, intra-cortical fiber myelination, etc. We used a clustering approach to group voxels with similar properties, based on relaxometry (T1) and DTI diffusivity (AD) characteristics (Figure 1). The resulting maps showed different microstructural patterns between cortical regions at the individual level, as well as strong progression according to the infants' age. This confirmed the early maturation of primary sensorimotor regions, followed by adjacent unimodal associative regions, and finally by higher-order transmodal associative regions. Nevertheless, this study should be reproduced at different ages, especially after one year of age, when the developmental patterns observed with T1 and with the fraction of water related to myelin still vary between primary and associative cortical regions (Deoni et al., 2015).

Multi-parametric MRI has also been used to explore the maturation of central grey nuclei, where intense proliferation of glial cells and membranes, and fiber myelination are observed over the preterm period and infancy. While univariate studies have highlighted marked microstructural changes with age, such as decrease in T1, T2, DTI diffusivities, and increase in anisotropy (Dubois, Alison et al. 2020), recent comparisons between multiple parameters have allowed to disentangle between maturational mechanisms (e.g. concentration of myelin-associated macromolecules, water content). In premature infants, measuring T1, DTI and MTR were relevant to fully characterize the microstructural properties and the maturational patterns of the globus pallidus, putamen, thalamus and ventolateral thalamic nucleus (Nossin-Manor, Card et al. 2013). For instance, this latter structure is already myelinated in the preterm period, which was confirmed by low T1 values, low DTI MD and RD diffusivities and high MTR, suggesting a high concentration of myelin-associated macromolecules, a low water content with high restriction, but this structure showed no specific directionality and coherence (low DTI anisotropy and AD) contrarily to white matter structures. Another study, combining T2, NODDI measures and the fraction of water related to the myelin, further highlighted that microstructural changes observed in the thalamus during the preterm period are not solely due to myelination (Melbourne, Eaton-Rosen et al. 2016).

### 3. *Quantifying white matter maturation*

Although some studies have suggested developmental relationships such as inverse correlations between MTR and T1 or T2 (Dubois, Dehaene-Lambertz et al. 2014, Dubois, Alison et al. 2020), different MRI parameters seem highly relevant to capture different properties of white matter maturation, which encourages combining or integrating these parameters. In a cuprizone mouse model of demyelination, a study has shown that the bound pool fraction from 'magnetization transfer imaging' (MTI) is the best indicator of the myelin sheath fraction, while T1 relates to the fraction of myelinated axons, and DTI AD to the fraction of non-myelinated cells (Thiessen, Zhang et al. 2013). Compared with T1 and T2, the fraction of water related to myelin also seems to provide complementary sensitivity to the white matter maturation (Deoni, Dean et al. 2012, Chen, Chen et al. 2019).

On the one hand, multi-parametric studies have been considered on the basis of diffusion MRI only, with the hypothesis that sequential changes in DTI parameters rely on successive steps of white matter maturation (Dubois, Dehaene-Lambertz et al. 2008, Dubois, Dehaene-Lambertz et al. 2014, Nossin-Manor, Card et al. 2015, Ouyang, Dubois et al. 2019). First, early changes in microstructure related to the fibers pre-myelination would mainly lead to a decrease in AD and RD. Subsequent wrapping of myelin sheaths around axons would not modify axial diffusivity but would decrease radial diffusivity, implying an additional increase in anisotropy. With such a DTI model, we were able to identify relevant differences in maturation between white matter bundles in infants (Dubois, Dehaene-Lambertz et al. 2008). With a clustering approach based on DTI parameters, we showed different microstructural characteristics between bundles of the language network, as well as maturational asynchrony (Dubois, Poupon et al. 2016): the ventral pathways (uncinate, fronto-occipital, middle and inferior longitudinal fascicles) appeared more mature than the dorsal ones (arcuate and superior longitudinal fascicles), although this difference decreased during infancy. Maturational differences between white matter bundles were also highlighted between birth and 2 years of age by combining the age trajectories of DTI parameters, and evaluating their asymptotes, delays and speeds (Sadeghi, Prastawa et al. 2013, Sadeghi, Gilmore et al. 2017). Beyond DTI, NODDI studies in newborns and infants have shown differences between white matter regions in terms of intra-neurite volume fraction (informing on the axonal fibers maturation) and orientation dispersion index (reflecting the presence of fiber crossings and fanning) (Kunz, Zhang et al. 2014, Jelescu, Veraart et al. 2015, Dean, Planalp et al. 2017). Thus comparing these parameters may help distinguish between bundles with similar cellular structure but different myelination (e.g posterior vs anterior limb of the internal capsule), or reciprocally with similar maturation but different fiber microstructural organization (e.g. external capsule vs periventricular crossroads) (Kunz, Zhang et al. 2014).

Diffusion parameters were also combined with other quantitative parameters to highlight maturation patterns of white matter regions, notably in premature infants. The comparison of T1, DTI parameters and MTR showed the high organization and packing of corpus callosum fibers during the preterm period despite their low myelination (Nossin-Manor, Card et al. 2013). And voxel-wise analyses further highlighted the lamination pattern in the cerebral wall, as well as different maturation mechanisms in the brain compartments (e.g. intermediate zone, subplate) (Nossin-Manor, Card et al. 2015). In the posterior white matter, T2 and NODDI changes were related over the same period and early infancy, but were not related to the fraction of water related to myelin, which suggested that the two former parameters rely on axonal and glial proliferation rather than on the myelin water content (Melbourne, Eaton-Rosen et al. 2016).

In a recent study of typical infants, we also proposed to combine T1, T2 and DTI diffusivities in white matter bundles, to provide an original measure of maturation based on the computation of the Mahalanobis distance comparing infants' individual data with a group of adults, while taking into account the possible correlations between MRI measurements (Kulikova, Hertz-Pannier et al. 2015). This revealed more maturational relationships between bundles than univariate approaches, and it allowed us to quantify their relative delays of maturation. The results confirmed the intense changes during the first post-natal year, as well as the maturational asynchrony, notably with early maturation of the spino-thalamic tract, optic radiations, cortico-spinal tract and fornix, and delayed maturation of associative bundles such as the superior longitudinal and arcuate fasciculi (Figure 2). In the same group of infants, the clustering approach of voxels described in previous section for the cortex led us to uncover four white matter regions with different compactness, maturation and anatomically relevant spatial distribution (Lebenberg, Poupon et al. 2015).

The last multi-parametric method that can be mentioned concerning the assessment of white matter maturation is the one that aims to quantify the so called "g-ratio" (i.e. the ratio of the axon diameter to the outer fiber diameter including the myelin sheath) (Stikov, Campbell et al. 2015). This composite marker can be estimated using diffusion MRI data (e.g. NODDI indices) and MTR or the fraction of water related to myelin. Studies in premature infants (Melbourne, Eaton-Rosen et al. 2014) and young children (Dean, O'Muircheartaigh et al. 2016) have shown that this marker decreases with age in the white matter, because of the myelination process. This index might provide relevant information on the efficiency of neural information transfer and on the conduction velocity of white matter pathways, but it remains difficult to estimate and interpret in the developing brain (Dubois, Adibpour et al. 2016).

Therefore, these studies in newborns and infants have shown that multi-parametric MRI approaches provide a more accurate description of maturation patterns in the developing brain, by considering complementary mechanisms through the white matter, and also the cortical and deep grey regions. Multi-modal MRI can also be used to relate different aspects of structural development, including morphological and microstructural properties of the brain tissues.

### III. Multi-modal MRI to relate complementary aspects of structural development

#### 1. *Relating morphometry, morphology and microstructure during early development*

During the last trimester of pregnancy and infancy, the cortical surface area dramatically increases (from ~150cm$^2$ at 27w PMA (Makropoulos, Aljabar et al. 2016) to ~700cm$^2$ and

~2000cm$^2$ at 1 and 24 post-natal months respectively (Lyall, Shi et al. 2015)), and the brain morphology becomes increasingly complex with the formation of gyri, primary, secondary and tertiary sulci (Dubois, Lefevre et al. 2019). Although underlying mechanisms are still widely discussed (Welker 1990, Llinares-Benadero and Borrell 2019), this folding process might support the enormous increase in cortical volume while maintaining reasonable connection distances (and thus information transmission times) between brain regions. So these intense macroscopic changes are probably the visible markers of changes in the microstructure of the cortical plate (the future cortex), over pre-term and early post-term periods marked by synaptic outburst and pruning, modifications in dendritic branching and fiber myelination (Kostović, Sedmak et al. 2019).

Recent multi-modal MRI studies, combining anatomical and diffusion MRI, have aimed to relate these morphological and microstructural features in the newborn brain. It was first proposed to compute a "radiality index" measuring the local directional coherence between the direction normal to the cortical surface and the main orientation of diffusion in the cortex, estimated with DTI model or 'distribution of 3D anisotropic microstructural environments in diffusion-compartment imaging' (DIAMOND) (Eaton-Rosen, Scherrer et al. 2017). This showed strong age-related decrease during the preterm period, as well as fastest changes in the occipital lobe and slowest changes in the frontal and temporal lobes. This decrease in radiality index paralleled the decreases in diffusion anisotropy and MD observed in the cortex, but correlations between these parameters were not evaluated beyond the effects of age. Note that this study preferred not to include sulcal regions in the analysis for methodological reasons, while previous observations described more intense age-related DTI changes in gyri than in sulci (Ball, Srinivasan et al. 2013, Ouyang, Dubois et al. 2019). Another study in premature infants compared changes in cortical volume, surface curvature, and microstructural characteristics of the cortex as measured with DTI and NODDI parameters (Batalle, O'Muircheartaigh et al. 2019). While important correlations were reported between these morphological and microstructural measures mainly during the period between 25 and 38w PMA and to a lesser extent during the 38-47w period, no correlations between cortical volume and diffusion parameters were observed when age dependencies were taken into account. However, some partial correlations were observed between mean curvature and diffusion parameters in several cortical regions: negative correlations with DTI anisotropy, as well as positive correlations with NODDI orientation dispersion index.

These studies suggested that changes in cortical microstructure might be at least partially related to the gyrification process. To test this hypothesis, we investigated whether different stages of microstructural maturation could be detected in cortical regions that fold successively, in preterm infants imaged longitudinally at around 30 and 40w PMA (Hertz, Pepe et al. 2018). We used anatomical MRI to perform a spectral analysis of gyrification (SPANGY) allowing us to detail the spatial-frequency structure of cortical patterns (Dubois, Lefevre et al. 2019), and we combined these measures with DTI information at the two ages (Figure 3). We first highlighted that the proxies of primary folds had an advanced microstructural maturation at 30w PMA (i.e. lowest DTI anisotropy and AD values) (Hertz, Pepe et al. 2018). Furthermore, the progression until term-equivalent age was lowest in these already well-developed regions, whereas it was most intense in the secondary fold proxies which grew considerably during this period. These findings were in agreement with a recent study in developing macaques which showed that cortical structural differentiation is coupled to the sulci formation rather than surface expansion (Wang, Studholme et al. 2017). Nevertheless, this still did not allow to decipher whether the folding process induces microstructural maturation, or conversely if regions with an advanced microstructure become folded.

Few other studies have aimed to compare cortical thickness and microstructural properties during early brain development. From 1 to 6 years of age, there is almost no relationship between the thickness and the intra-cortical fraction of water related to myelin (Deoni, Dean et al. 2015). In infants in the first post-natal year, estimating changes in cortical thickness with MRI is difficult because the contrast between grey and white matter evolves on T1w and T2w images due to myelination of intra- and subcortical fibers. A dedicated pipeline of longitudinal data was recently used to characterize the temporal evolution of cortical thickness from 1 to 24 months of age (Wang, Lian et al. 2019). Nevertheless, the interpretation of thickness changes is subject to discussion given that this MRI measure is sensitive to the subject's motion and to the ongoing maturation of tissues (Ducharme, Albaugh et al. 2016, Walhovd, Fjell et al. 2017, Dubois, Alison et al. 2020). Therefore, it would be informative to couple this information with measurements of the cortical microstructure.

## 2. *Considering morphological patterns to assess brain tissue maturation*

In order to study the maturation of tissues, especially the cortex, reliably between infants, it is necessary to consider similar regions between individuals, as adjacent regions may be at different stages of maturation. This is even more critical when a voxel-wise method is used, rather than a method based on regions of interest. And this is particularly challenging for studies of premature newborns and infants because the brain shows significant age-related growth in size and folding. Aligning brains in a common space through procedures of registration and spatial normalization therefore requires specific methods in this population of subjects (Dubois, Alison et al. 2020).

Recently, a 2-step landmark-based strategy allowed us to register the brains of preterm newborns, infants and various databases of adults (Lebenberg, Labit et al. 2018). The DISCO method (for 'diffeomorphic sulcal-based cortical registration') was used to embed sulcal constraints in a registration framework used to initialize the DARTEL step (for 'diffeomorphic anatomical registration using exponentiated Lie algebra') which provided an accurate registration of cortical ribbons. This methodology allowed us to project an individual parcellation of cortical regions (Kabdebon, Leroy et al. 2014) on a group of newborns. By coupling this anatomical framework with a precise mapping of diffusion parameters (DTI and NODDI), we were able to detail the developing microstructure of several motor and sensory regions (Chauvel, Rheault et al. 2020). We highlighted important differences along the pre- and post-central gyri, which might be related to the progressive functional specialization for the different parts of the body.

Comparing the microstructure of similar cortical regions across the left and right hemispheres also requires taking into account the asymmetries that are observed from an early age in the shape and folding of the hemispheres (e.g. at the level of the petalias, the sylvian fissure, the superior temporal sulcus) (Dubois and Dehaene-Lambertz 2015, de Vareilles, Rivière et al. 2020). Using a similar DISCO-DARTEL framework to register left and right hemispheres of infants aged 1-5 months, we were able to compensate for morphological asymmetries and evaluate microstructural asymmetries based on voxel-wise analyses of DTI cortical maps (Rolland, Lebenberg et al. 2019). This study highlighted asymmetrical microstructural organization in specific sensorimotor and language regions of infants, suggesting a structural basis for functional lateralization.

## 3. *Linking the maturation patterns of different brain tissues*

Previous studies have shown the value of linking anatomical MRI and quantitative MRI modalities such as diffusion MRI to better understand some aspects of brain development. Other multi-modal studies have aimed to link the maturation of different tissues. As detailed in part II, spatial gradients of maturation have been described in the grey and white matter. But these changes are not taking place independently since synchronous development between cortical and adjacent white matter regions were observed in primary motor, primary visual, visual association and prefrontal regions (Smyser, Smyser et al. 2016). This again suggested that maturation in the primary motor and sensory regions precedes maturation in the association areas. Another study showed that the development of thalamic substructures around term age is synchronized with the maturation of their respective thalamo-cortical connections, to the frontal, precentral, postcentral, temporal and parieto-occipital cortices (Poh, Li et al. 2015). More generally, the advancement of grey and white matter maturation seemed inter-related and dependent on the underlying brain connectivity architecture, as corresponding maturation levels were found in cortical regions and their incident connections in newborns, and also in connected regions (Friedrichs-Maeder, Griffa et al. 2017). From 1 to 6 years of age, regional measures of cortical thickness were found to be partially driven by changes in adjacent white matter myelination, suggesting that cortical and white matter maturation reflect distinct, but complimentary, neurodevelopmental processes (Croteau-Chonka, Dean et al. 2016). In older children (5-18 years old), covariation patterns were observed for different cortical shape measures and subcortical grey matter volumes, and considering this multi-modal information provided an accurate prediction of a person's age, sex and general cognitive ability (Zhao, Klein et al. 2019). But this type of study remains to be done in the infant's brain. Finally, multi-modal studies go beyond exploring the structural development of the brain, linking it to its functional development.

## IV. Multi-modal neuroimaging to compare structural and functional brain development

### 1. *Relating functional specialization, morphometry and microstructure*

To date, the relationship between functional specialization and structural properties of the brain has been little explored during development. Yet it is an important question, as some recent studies in adults are beginning to show, particularly for the sensorimotor network. Following on from the observations made for the hand knob, studies combining anatomical MRI and task-based fMRI have outlined that functional representations of distinct parts of the body in the primary motor and somatosensory cortex (somatotopic arrangement) have a precise spatial correspondence with the morphological features of the regions, as shown for the sulcal segments of the central sulcus (Germann, Chakravarty et al. 2019) and postcentral sulcus (Zlatkina, Amiez et al. 2016). For instance, the transverse postcentral sulcus, when present, appeared to be functionally related to the oral sensorimotor representation (mouth and tongue). This indicates that exploring the sulcal morphology may inform functional specialization. Functional representations of body parts have also been linked to patterns of cortical myelination and to the topography of functional connections (connectopies) measured with T1w/T2w ratio and rs-fMRI respectively (Kuehn, Dinse et al. 2017, Haak, Marquand et al. 2018). These studies raise fundamental questions regarding the interplay between functional specialization, morphological growth, development of connections and maturation of cortical microstructure. Further studies in newborns and infants are needed to investigate this complex issue in the future.

### 2. *Linking structural and functional connectivity patterns*

In the recent years, the combined evaluation of structural and functional connectivity has been made possible by the acquisition of diffusion MRI and rs-fMRI data in the same individuals. A recent study in children and adolescents has suggested that the coupling between connectivity patterns might support the development of functional specialization and cognition (Baum, Cui et al. 2020). However it is important to note that the mechanisms linking these measures are still poorly understood even in adults (Smyser, Snyder et al. 2011): the strength of structural connections seem to predict the strength of functional connections, whereas strong functional connections exist between regions with no direct structural connections. And some have suggested that structural approaches could primarily reflect monosynaptic connections while functional approaches could also be sensitive to polysynaptic connections.

In the developing brain, a few multi-modal studies have reported hierarchical structural maturation from primary to higher-order cortices, which is partially paralleled by functional development (Cao, Huang et al. 2017). In newborns, the similarities and dissimilarities between structural and functional connectivity patterns depend on the systems, with greater overlap in primary sensory networks than in higher-order transmodal ones where divergence in spatial patterns are observed (van den Heuvel, Kersbergen et al. 2015, Ferradal, Gagoski et al. 2019, Larivière, Vos de Wael et al. 2019). This might be due to differences in maturation between systems, leading to technical biases in diffusion MRI (i.e. more reliable tractography reconstructions of mature structural connections) (Dubois, Adibpour et al. 2016). Others have suggested that the structural network might remain ahead and pave the way for the development of the functional brain network (Zhao, Xu et al. 2018). Despite already established structural connections, associative regions might show delayed functional integration and segregation, and this might contribute to the observation of increased long-range functional connectivity and decreased short-range connectivity during development (Fair, Cohen et al. 2009, Ouyang, Kang et al. 2017). Although still incomplete, overall these studies suggested that the broad outlines of the brain network architecture are in place at an early stage.

### 3. *Relating white matter maturation and EEG responses*

While several MRI modalities have detailed the progressive maturation of white matter networks, the functional efficiency of neural communication has been evaluated with EEG and MEG recordings for sensory modalities. These techniques enable to measure the latency of evoked responses (i.e. the averaged responses over multiple trials following successive stimulations) and show drastic decreases in response latencies during development (Dubois, Adibpour et al. 2016). Although the role played by of cortical maturation and synaptogenesis cannot be overlooked, white matter myelination is one of the key mechanisms that decrease the response latency, as it is known to significantly increase the conduction velocity of nerve impulses along axonal fibers (Baumann and Pham-Dinh 2001). At constant pathway length, it leads to a decrease in response latency during development. Conversely, as brain size increases with age, mainly during the first two years, it may be necessary to further increase conduction velocity by prolonging myelination just to maintain constant latency (Salami, Itami et al. 2003). As with structural changes, these functional changes occur at different times and speeds depending on cerebral regions and functions involved (Dubois, Adibpour et al. 2016). A few recent studies have sought to link decreased latency, increased conduction, and fibers myelination in a network by combining investigations with complementary MRI and EEG/MEG techniques in the same children. Again, the assessment of reliable anatomo-functional relationships required consideration of the age of the infants, which is the main factor explaining the developmental changes.

A few recent studies have investigated this issue for the visual modality that develops intensively after birth. Successive EEG/MEG visual evoked responses are recorded in occipital regions. At term birth, the first EEG positive component P1 (~P100 in adults) is detected at a latency that decreases strongly and quickly with age (Taylor, Menzies et al. 1987, Harding, Grose et al. 1989, McCulloch and Skarf 1991), from around 260ms in neonates to around 110-120ms at 12-14 weeks of age, depending on the patterns size (McCulloch, Orbach et al. 1999). In infants, we related the increase in P1 conduction speed to the maturation of optic radiations (the lower the radial diffusivity measured with DTI, the higher the speed) (Dubois, Dehaene-Lambertz et al. 2008) (Figure 4a). This relation was specific (i.e. not observed for other white matter bundles) and not explained by intra-individual differences in infants' age. Recently, this observation has been further extended to cortico-cortical connections. When visual stimuli were presented laterally (i.e. in a single hemifield), visual responses were first observed in the contralateral hemisphere, then in the ipsilateral hemisphere of infants. And the speed of the inter-hemispheric transfer of responses was related to the maturation of visual callosal fibers connecting the occipital regions (Adibpour, Dubois et al. 2018) (Figure 4b).

These anatomo-functional relationships have seemed less clear for the auditory system, suggesting more complex interaction mechanisms in which the environment may play a more important role. Indeed, the auditory modality is already functional *in utero*, but its development is more prolonged during early childhood than the visual and somatosensory modalities. Evoked responses show extended developmental changes throughout the early post-natal years (Dubois, Adibpour et al. 2016), which may make comparison with white matter maturation in infants more difficult. Recently, we have observed early P2 responses in infants following monaural auditory stimulations (i.e. in one ear at a time), both on the contralateral and ipsilateral sides of the brain. Response latencies decreased with age, and ipsilateral responses were significantly longer in the left hemisphere than in the right (Adibpour, Lebenberg et al. 2020), which was not the case in infants with agenesis of the corpus callosum (Adibpour, Dubois et al. 2018). These results suggested that left ipsilateral responses might include a transfer of right contralateral responses via the callosal fibers, while the reverse (left-to-right transfer) would not be observed. We further related the speed of left ipsilateral responses to the microstructure of the auditory callosal fibers connecting the temporal regions of the two hemispheres (Adibpour, Lebenberg et al. 2020) (Figure 4c). Such functional asymmetries relying on callosal fibers could influence the emergence of early lateralization of the language network and reinforce an initial bias during development.

To date, no studies have linked anatomical and functional maturation in the somatosensory modality, despite intense changes in electrophysiological responses during the preterm period and infancy (Dubois, Adibpour et al. 2016). Additional studies are therefore needed to confirm the findings for the visual modality in infants. And comparison of anatomo-functional changes between modalities in the same subjects would make it possible to characterize the asynchronous development of brain networks and explore their sensitivity to distinct critical periods and various environmental stimulations. In addition to the latency of evoked responses, several other EEG/MEG parameters (e.g. evolution in peaks morphology and amplitude, complexity measures) could be compared to the brain anatomical changes during development (Dubois, Adibpour et al. 2016).

### 4. *Comparing fMRI and EEG responses*

Since EEG and functional MRI do not reflect the same mechanisms (neural vs. hemodynamic activity), it is important to compare the measures and responses provided by these methods in order to better understand their changes and significance during development. Nevertheless, to our knowledge, only two multi-modal studies have used both EEG and fMRI in newborns so far. The first one considered simultaneous EEG-fMRI in preterm infants aged 32-36w PMA to localize the source of spontaneous neuronal bursts that are critical for brain maturation (the so-called "delta brush") (Arichi, Whitehead et al. 2017). It revealed that the insula, a densely connected hub of the developing brain, is a major source of transient bursting events in both the left and right hemispheres. It is interesting to note that this spontaneous activity, measured in preterm newborns shortly after birth, may be a good marker of brain development, as increased activity has been linked to more rapid growth of the brain and subcortical grey matter up to term-equivalent age (Benders, Palmu et al. 2015).

The second study investigated voice perception in newborns using two independent paradigms with high-density EEG and fMRI (Adam-Darque, Pittet et al. 2020). Results from both modalities suggested that the main components of the adult voice-processing networks are present early on and that preterm infants at term-equivalent age have enhanced processing for voices than full-term newborns. More studies are needed in the future to better understand how maturational changes in responses measured in EEG and fMRI are related during infancy.

## Conclusion and perspectives

In recent years, a wide variety of new MRI methods have been proposed and implemented to allow precise exploration of the developing brain in newborns and infants, targeting multiple mechanisms ranging from morphological to microstructural changes in grey and white matter, in addition to metabolic and functional changes. Nevertheless, such studies remain limited in several aspects due to inherent methodological and experimental challenges. Research is still needed to improve the reproducibility of acquired data and quality control. A major issue in the coming years will be to link the different scales and facets of developmental processes, and to relate molecular, cellular and network changes in a comprehensive integrative model to be compared with the cognitive development of infants. Multimodal MRI has a key role to play in this regard, as well as to better characterize early deviations from neurodevelopmental trajectories due to pre- or perinatal disturbances. Because MRI examinations can be performed well before the child's behavior and clinical outcome are known, it is an essential investigative method for evaluating the efficiency of early neuroprotective interventions or remediation strategies to avoid long-term disability in children. Of course, the diverse information provided by this technique is complementary to the many other factors responsible for inter-individual variability, such as intrauterine growth, gestational age at birth, socio-economic status, etc. For this reason, the multi-modality of research and clinical studies should far exceed that of MRI brain exploration, by including complementary neurophysiological, behavioral and clinical evaluations.

## Acknowledgments

The research was supported by grants from the Médisite Foundation (2018), the Fondation de France (call Neurodevelopment 2012), the Fyssen Foundation (2009), the European Union's Horizon 2020 Research and Innovation Programme (HBP 2013). This study contributed to the IdEx Université de Paris (ANR-18-IDEX-0001).

# Figures

*Figure 1: Multi-modal MRI to quantify the maturation of cortical regions.*

a: T1 and DTI axial diffusivity (AD) vary between brain regions according to their different maturation and microstructure patterns, as shown here on average maps obtained from a group of typical infants aged 1-5 months. Higher maturation and a more complex microstructure correspond to lower T1 and AD values.

b: A clustering approach based on a 'gaussian mixture model' (GMM) algorithm applied to T1 and AD values of all infants allowed us to identify 5 clusters of cortical voxels. Although we were not able to compare these results with equivalent measurements obtained in the mature brain, we could extrapolate the microstructure properties and the maturation order of these clusters according to T1 and AD values in infants, from the least mature (highest values, in blue) to the most mature (lowest values, in red).

c: The spatial distribution of these clusters was highly meaningful. While the individual maps showed that immature clusters were observed in the youngest infants, and conversely for mature clusters (not shown), the average maps on the infant group highlighted the heterogeneities in maturation and microstructure in the cortical regions. The most mature clusters (in red) were localized in the primary regions (around the central sulcus for sensorimotor regions, in Heschl gyrus for auditory regions, around the calcarine fissure for visual regions), whereas the intermediate (in yellow-green) and the least mature clusters (in green-blue) were observed in unimodal and transmodal associative regions respectively.

Adapted from (Lebenberg, Mangin et al. 2019).

### a. Average maps of quantitative parameters over the cortex of infants

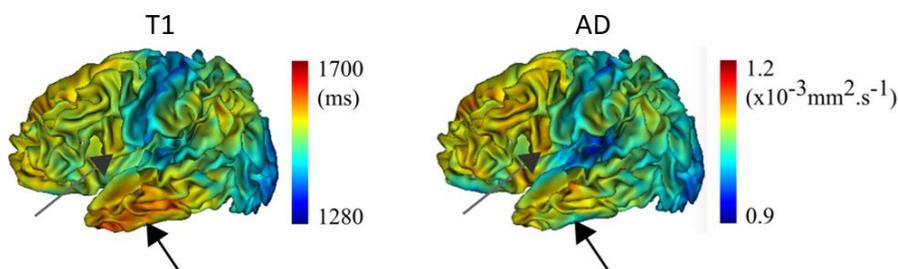

### b. Individual parameters in each cortical cluster

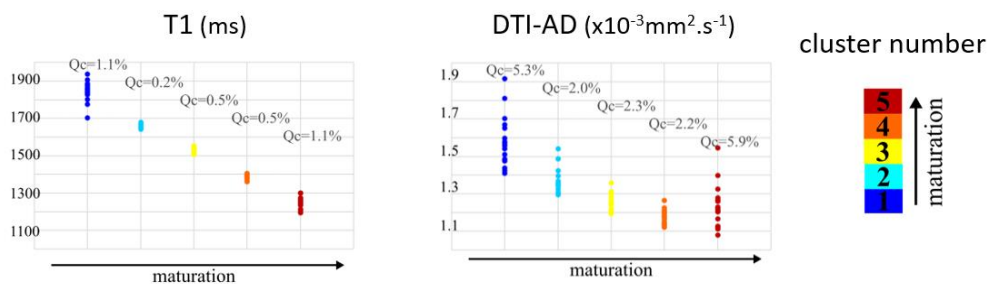

### c. Average map of the clusters identified in the cortex

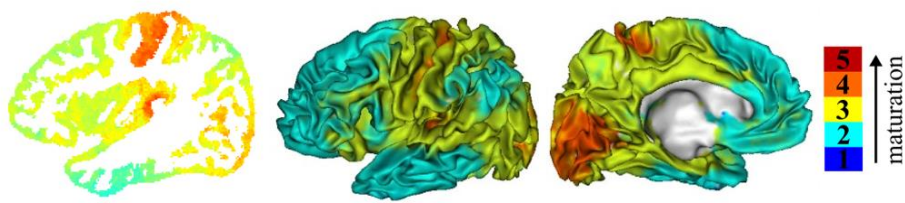

*Figure 2: Multi-modal MRI to quantify the maturation of white matter bundles.*

a: Maps of quantitative parameters (T1 and T2 relaxation times, DTI diffusivities: axial AD and radial RD) show different contrasts and developmental changes in the white matter, as illustrated here for a 2-month old infant and an adult.

b: Strong decreases of these four parameters are observed in the white matter between 1 and 5 months of age (the lines show significant correlations with age). This is illustrated for bundles of different functional networks: projection bundles like the optic radiations (OR) and the inferior portion of the cortico-spinal tract (CST), limbic bundles like the inferior branch of the cingulum (CG), associative bundles like the external capsule (EC) and arcuate fasciculus (AF). The lower the parameters are, the more the bundle has a complex microstructure (with high compactness or myelination) or the more mature it is. But the order between bundles is not reproducible between parameters.

c: A multi-parametric approach combining these four parameters in each infant compared to a group of adults allowed us to estimate the maturation of white matter bundles in a more reliable way. The resulting Mahalanobis distance decreases with the infants' age and shows strong heterogeneities in maturation across projection, callosal, limbic and association bundles (the color codes for maturation: from blue in the least mature bundles to red in the most mature ones).

Adapted from (Dubois, Dehaene-Lambertz et al. 2014, Dubois, Kostovic et al. 2015, Kulikova, Hertz-Pannier et al. 2015).

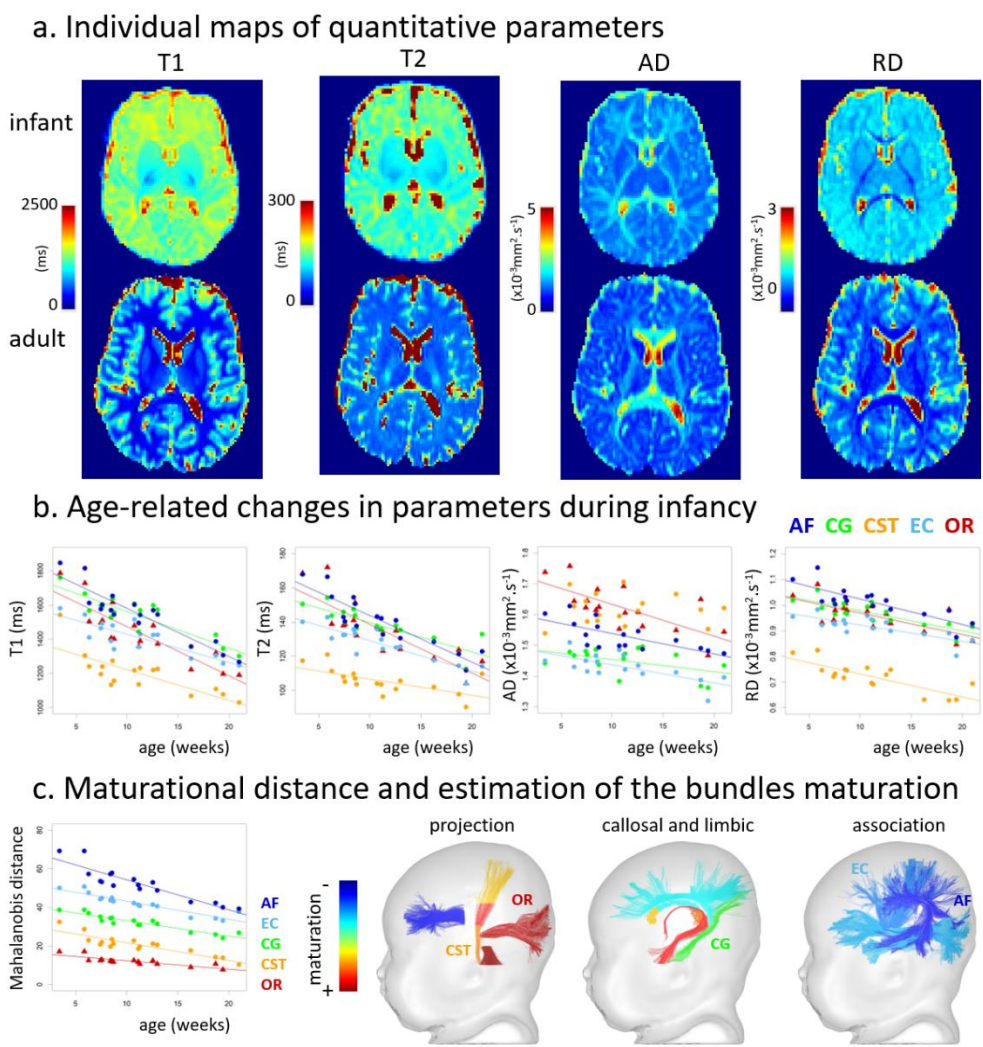

*Figure 3: Multi-modal MRI to relate the folding process and cortical microstructure.*

a: In preterm newborns imaged longitudinally at around 30w PMA [29-32w PMA] and at term equivalent age [40-42w PMA], individual registration of inner cortical surfaces at the two ages was performed using a spectral-based algorithm with global surface matching.

b: Spectral analysis of gyrification (SPANGY) on surface curvature was used to identify proxies of the developmentally defined parcels (primary, secondary and tertiary folds) at the two ages.

c: DTI color-coded directionality map showed early radial organization of the cortex at ~30w PMA (arrows) but not at term equivalent age. We observed a strong decrease in the DTI parameters measured in the cortical ribbon (anisotropy, axial diffusivity AD) with the age of the infants.

d: ANCOVA-Tukey analyses were performed to evaluate differences in DTI parameters between SPANGY parcels (red bars: $p<0.001$ after correction for multiple comparisons). Proxies of primary folds (in dark blue) showed the lowest DTI parameters (AD shown here) at ~30w PMA and the smallest changes up to term-equivalent age, suggesting an early complex microstructure and an advanced maturation pattern. Secondary folds (in light blue), and tertiary folds to a lesser extent (in green), showed lower DTI indices than gyri and the most significant changes between ~30 and ~40w PMA, suggesting that regions that are folding have a changing microstructure.

Adapted from (Hertz, Pepe et al. 2018).

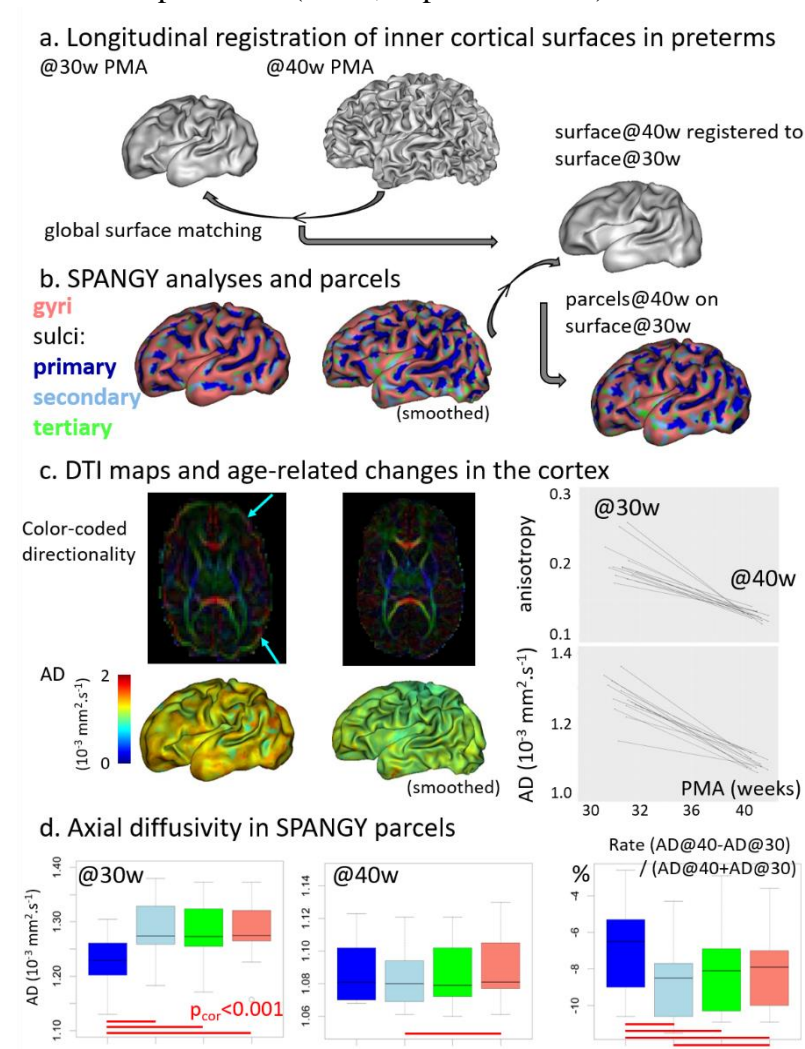

*Figure 4: Multi-modal neuroimaging to relate structural and functional development.*

a: During infancy, the speed of P1 responses to visual stimuli increases with age, while the visual pathways become myelinated, resulting in a decrease in DTI radial diffusivity in the optic radiations. For the visual system, these functional and structural markers of maturation have been related beyond age dependencies. Adapted from (Dubois, Dehaene-Lambertz et al. 2008).

b: Similar relationships have been observed for responses to visual stimuli presented laterally (in one hemifield at a time). The speed of the responses transfer, from the contralateral to the ipsilateral hemisphere, have been related to the maturation of callosal fibers connecting visual regions. Adapted from (Adibpour, Dubois et al. 2018).

c: Although such relationships are more difficult to demonstrate for the auditory system, we observed that the speed of ipsilateral responses measured in the left hemisphere following stimuli presented in the left ear, is related to the maturation of callosal fibers connecting auditory regions. This suggests that early structural biases might lead to the functional lateralization for speech processing in the left hemisphere. Adapted from (Adibpour, Lebenberg et al. 2020).

### a. Visual responses to central stimuli

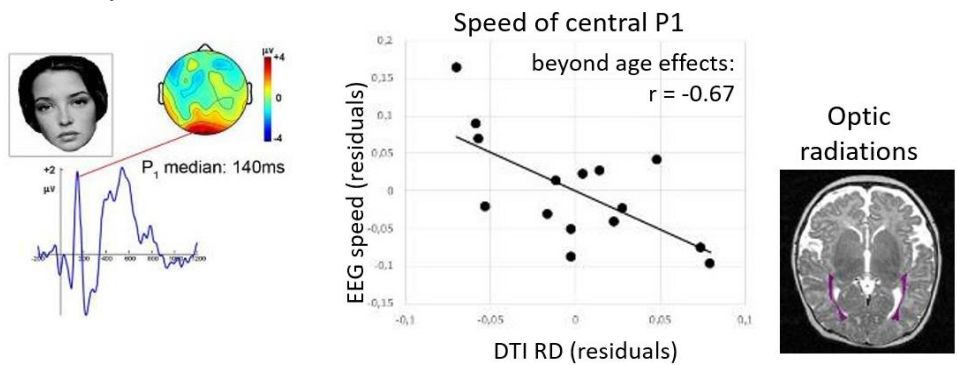

### b. Visual responses to lateral stimuli

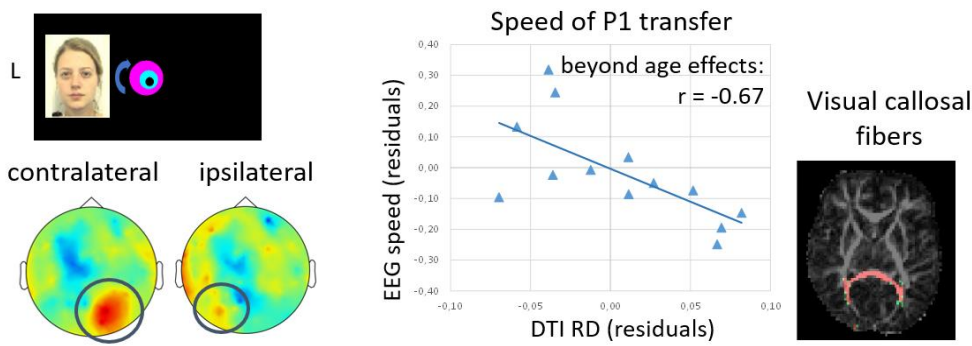

### c. Auditory responses to lateral stimuli

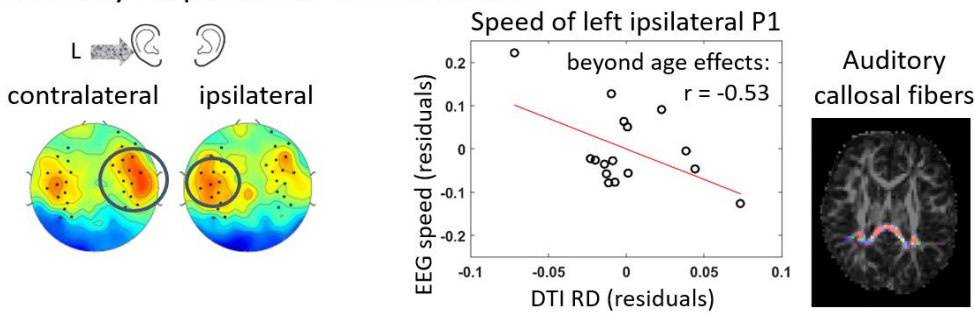